# The physics of stripe patterns in turbulent channel flow determined by DNS results

**P. KIŠ, Y. JIN, H. HERWIG**[*]

* Corresponding author
Institute for Thermo-Fluid Dynamics, Hamburg University of Technology,
Denickestr. 17, 21073 Hamburg, Germany

The turbulent flow in an infinitely extended plane channel is analysed by solving the Navier-Stokes equations with a DNS approach. Solutions are obtained in a numerical solution domain of finite size in the streamwise as well as in the lateral direction setting periodic boundary conditions in both directions. Their impact on large scale structures in the turbulent flow field is analysed carefully in order to avoid their suppression. When this is done appropriately well known stripe patterns in these flows can be observed and analysed especially with respect to their relative motion compared to the mean flow velocity. Various details of this stripe pattern dominated velocity field are shown. Also global parameters like the friction factor in the flow field and the Nusselt number in the temperature field are determined based on the statistics of the flow and temperature data in a very large time period that guarantees fully developed turbulent flow and heat transfer.

## 1. Introduction

Direct numerical simulation (DNS) is a powerful tool in turbulence research provided it is applied according to the specific requirements of the particular flow situation under consideration. In the Reynolds number range of transitional channel flow this implies that large scale structures have to be accounted for appropriately. This, however, is a challenging demand since the DNS of such a benchmark case requires periodic boundary conditions which, by their nature, interfere with the physics of turbulence.

In the low Reynolds number range large scale turbulent structures appear in different geometries and for various flow situations. In pipe flows, they are called slugs and puffs, associated with the transition from laminar to turbulent flow and an incomplete relaminarization process, respectively. They were first observed in transitional pipe flow by Wygnanski & Champagne (1973) and Wygnanski et al. (1975). In channel flow, what are puffs in pipe flow, appear as elongated near–wall streaks, which are smoothly staggered forming stripe patterns inclined at certain angles against the streamwise direction. The stripe itself consists of a center region which is highly turbulent and which is confined by well ordered streaks which are less turbulent and show a tendency towards laminar behaviour. The stripe patterns in a DNS can be very stable once the periodic boundary conditions support the formation of these structures. Their appearance could also be observed experimentally, see Hashimoto et al. (2009).

Various aspects of the large scale structures and especially their formation have been analysed by DNS, see for example Tsukahara et al.(2005) and Tsukahara et al.(2006). These large scale structures are confirmed by the experiments in a rectangular channel (Lemoult et al., 2013). The experiments conducted in a planar channel flow by Seki & Matsubara (2012) show that the large-scale fluctuation exits even beyond the critical Reynolds number 2600. Tuckerman et al. (2014) simulated the oblique stripes in plane Poiseuile flow in the transitional range with a DNS method. The results show the gradual development from uniform turbulence to oblique stripe patterns and finally to a laminar flow when the Reynolds number is reduced from 2300 to 800.

Related patterns have been found in various flow situations, like Couette flow (see e.g. Barkley & Tuckerman(2005), Duguet et al. (2010)), Taylor–Couette flow between counter–rotating cylinders (see e.g. Coles(1965) and Van Atta(1966)), and shear flow between counter–rotating disks (see e.g. Cros & Le Gal(2002)). With the plane Couette flow as an example, Duguet &

Schlatter (2012) show analytically that the corresponding laminar-turbulent stripes are always oblique with respect to the mean direction of the flow. Brethouwer, et al. (2012) simulated subcritical rotating, stratified and magneto-hydrodynamic wall-bounded flows in large computational domains by a DNS method. Their study indicates a regime of large-scale oblique laminar-turbulent patterns which can be observed up to large values of the Reynolds number Re when damping is increased by the Coriolis, buoyancy or Lorentz force. Philip & Manneville confirmed the existence of patterns made of alternately laminar and turbulent oblique bands in plane Couette flow. Based on the experimental studies, Manneville (2014) and Seshasayanan & Manneville (2015) suggested theoretical interpretations with respect to oblique stripes in the transitional regime of wall-bounded flows, particularly plane Couette flow.

The focus of our study will be on the physics of the fully developed stripe patterns. Therefore DNS results for a very large time span are required in order to deduce reliable time averaged statistical data from these results. Based on these data we will carefully analyse the flow field with respect to the movement of the stripe patterns. Prior to this, a comprehensive study of the adequate domain size is made when large scale structures occur. Also the temperature field will be considered including entropy generation considerations that can lead to a heat transfer correlation in terms of the Nusselt number.

The present study is a systematic investigation with respect to the physics of the large scale patterns in plane channel flow. Also the requirements are addressed that DNS solutions have to fulfill when these structures occur and have to be resolved appropriately. For both aspects we use our own DNS code which is described to some detail in the appendix of this paper. Further details and data for various flow situations gained by this code can be found at http://www.tu-harburg.de/tt/dnsdatabase/dbindex.de.html.

## 2. Channel flow and its Moody chart based on DNS results

An important overall parameter of a channel flow is its friction factor $f = f(Re_{Dh}, K)$ with f being a function of the Reynolds number $Re_{Dh} = \rho u_m 4\delta/\mu$ and the roughness number $K = k/4\delta$. The characteristic length here is the hydraulic diameter, which for a plane channel is four times the channel half width $\delta$. Often, $f$ is defined in terms of a time mean pressure gradient $d\bar{p}/dx$ or a time mean wall shear stress $\bar{\tau}_w$ as

$$f \equiv -\frac{d\bar{p}}{dx}\frac{8\delta}{\rho\,u_m^2} = \frac{8\bar{\tau}_w}{\rho\,u_m^2} \tag{2.1}$$

This, however, is only a reasonable definition of f when the flow is fully developed and horizontal. A more general definition, including (2.1), is

$$f \equiv \frac{d\varphi}{dx}\frac{8\delta}{u_m^2} \tag{2.2}$$

with $\varphi$ as time–averaged specific dissipation in a cross section $x$ = const, see Herwig et al. (2008) for details. With this definition f is linked to the dissipation process and thus also to the entropy generation that always occurs when there are losses (of exergy or available work) in a flow field. More details about this approach which is also called second law analysis (SLA) can be found in Herwig (2011), Herwig & Wenterodt (2011) and Kis & Herwig (2011) when applied to DNS.
Figure 1 shows the two parallel walls between which the channel is formed, together with the coordinate system and the solution domain of size $L_x \times 2\delta \times L_z$. With the mean flow in x-direction there is a pressure gradient $dp/dx < 0$ and a wall shear stress

$$\bar{\tau}_w = \frac{1}{2}\mu\left(\left.\frac{\partial \bar{u}}{\partial y}\right|_{y=-\delta} - \left.\frac{\partial \bar{u}}{\partial y}\right|_{y=\delta}\right) > 0 \tag{2.3}$$

which, for a fully developed horizontal flow, can both be used to determine $f$. Together with the specific dissipation φ according to (2.2) there are three independent ways to find f in such a flow, given as $f_p$, $f_w$ and $f_\varphi$, respectively. Deviations between these values for $f$ due to discretization errors are used in the appendix to assess the accuracy of the DNS solutions in their respective domains.

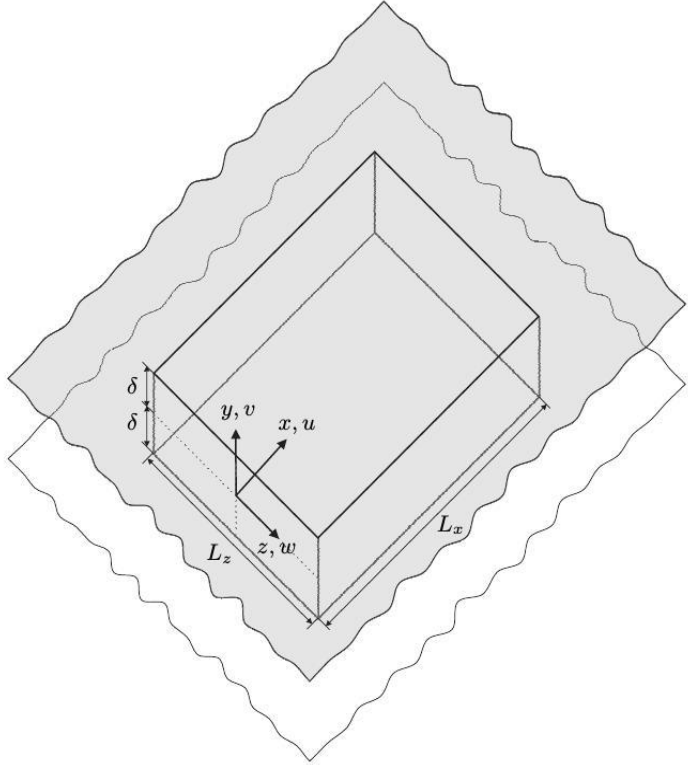

Figure 1: Plane channel between two parallel walls and the solution domain of size $L_x \times 2\delta \times L_z$; coordinates $x$, $y$, $z$ and the corresponding velocity components $u$, $v$, $w$

### 2.1. *Domain size considerations*

In this study we quite generally restrict ourselves to the case of smooth walls, i.e. with $K = 0$ in the friction law. For this case figure 2 shows a Moody chart with all results gained by DNS solutions with our DNS code in terms of $f_w$, i.e. $f$ determined from the wall shear stress. The size of the flow domain with periodic boundary conditions in $x$- and $z$-direction is crucial. It is different for different ranges of the Reynolds number and will be discussed in the following. For convenience, the extent of the solution domain in all three directions is specified by referring it to the channel half width $\delta$, with that in y-direction being 2 and that in the other two directions x and z being an integer multiple of $\pi$, like $5\pi$ and $2\pi$, for example. For the later discussion our results for the famous cases of Kim et al.(1987) and Moser et al.(1999) with a domain size of $4\pi \times 2 \times 2\pi$ and $2\pi \times 2 \times \pi$, respectively, are specially marked as MKM.

Starting at high Reynolds numbers $Re_{Dh} \approx 10^5$, it turns out that a domain size of $2\pi \times 2 \times \pi$ is quite sufficient with no noteworthing effects when this size is increased. All relevant scales of the turbulent structures are well resolved and the periodic boundary conditions set in *x*- and *z*-direction do not interfere with large scale structures. This can also be concluded from the time mean autocorrelation function $R_{uuX}$, for example, defined as

$$R_{uuX}(y,r) = \frac{1}{\Delta t\, L_z} \int_{t_0}^{t_0+\Delta t} \int_0^{L_z} u(0,y,z,t) u(r,y,z,t)\, dz dt - \overline{u}^2(y)$$
$$= \overline{u'(0,y) u'(r,y)} \qquad \text{with } 0 \leqslant r \leqslant L_x \tag{2.4}$$

Figure 3 shows this function at *y* = 0, i.e. in the channel mid–plane for $Re_{Dh}$ = 9333 and two different domain sizes $3\pi \times 2 \times 2\pi$ and $5\pi \times 2 \times 2\pi$, respectively. Starting from *r* = 0 the correlation is the same for both domains close to *r* = 0. Since, however, periodic boundary conditions are set at $r = L_x$, i.e. *r* = $3\pi$ and $5\pi$, respectively, they do not coincide further downstream. Figure 3 may give an impression of how a periodic boundary condition affects the flow field. It is generally assumed that this unphysical constraint has a negligible overall effect when periodic boundary conditions are introduced. The bigger the domain size the more accurate this assumption obviously is since then autocorrelation values are closer to zero in the central part of the solution domain.

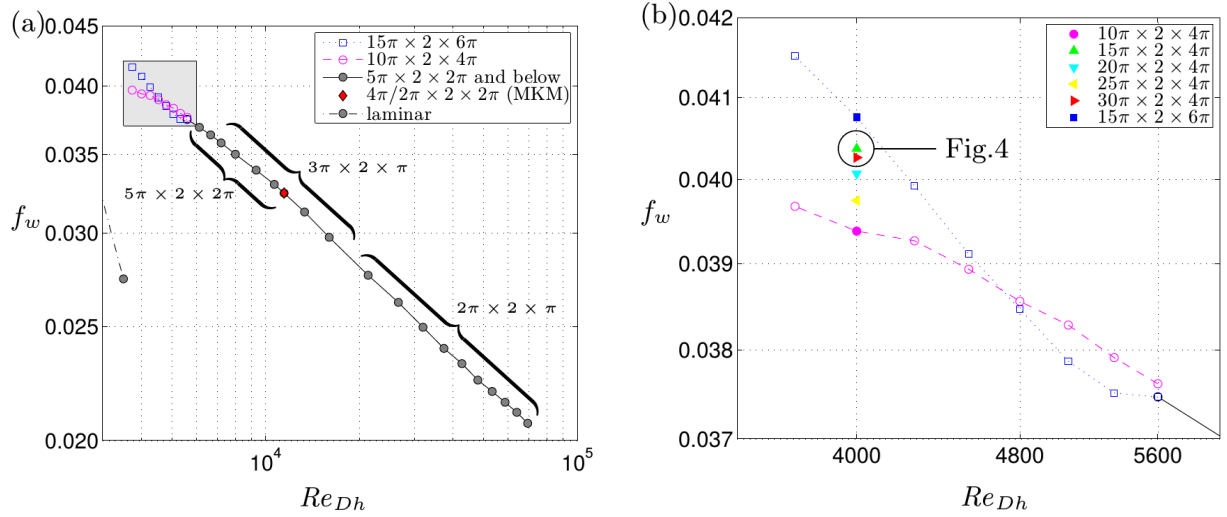

Figure 2: Moody chart for the plane channel with smooth walls ($K = 0$); results from DNS in solution domains with double periodic boundary conditions ($x$- and $z$-direction); domain size indicated by $L_x/\delta \times 2 \times L_z/\delta$, (a) large $Re$-range, (b) enlarged part marked in (a)

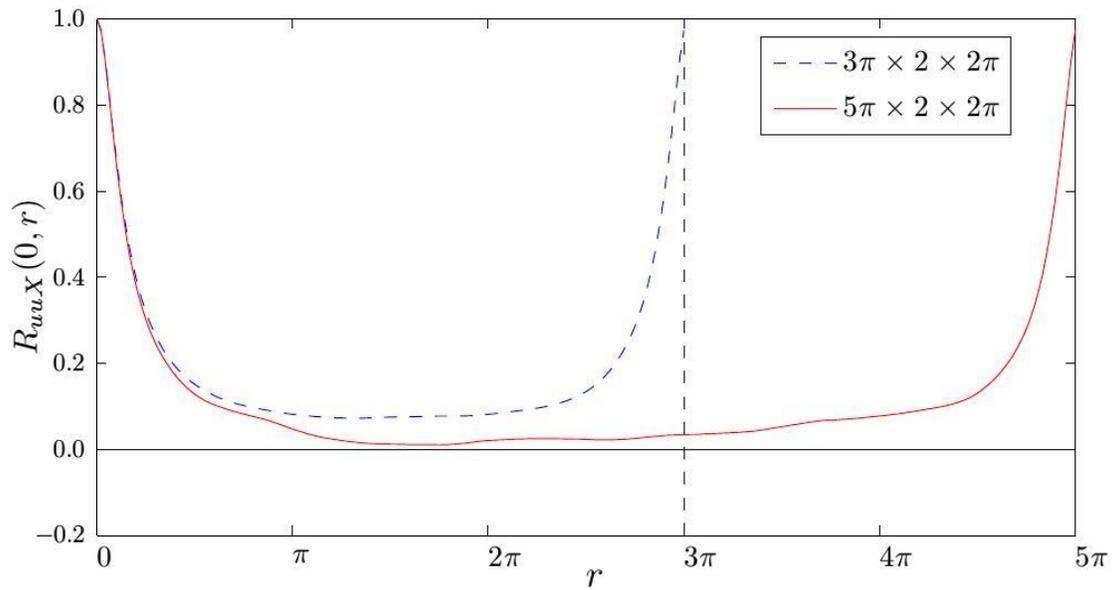

Figure 3: Autocorrelation function for two different domain sizes; Re Dh = 9333

When the Reynolds number is reduced below $Re_{Dh} \approx 6000$, which corresponds to $Re_\tau \approx 100$, a very different situation occurs. Here, $Re_\tau$ with $u_\tau = \sqrt{\overline{\tau}/\rho}$ is defined as

$$Re_\tau \equiv \frac{\rho\, u_\tau \delta}{\mu} = \sqrt{\frac{1}{4} Re_{Dh} \left|\frac{\partial \overline{u}}{\partial y}\right|_w} \qquad (2.5)$$

For these low Reynolds numbers the friction factor $f$ is strongly affected by the choice of $L_x$ and $L_z$, i.e. the size of the solution domain. This is due to the occurrence of large scale structures once the Reynolds number is reduced below $Re_\tau \approx 100$, for a plane channel flow. This will be discussed in detail with six different domain sizes for $Re_{Dh} = 4000$, corresponding to $Re_\tau \approx 70$, which are indicated in figure 2 (b).

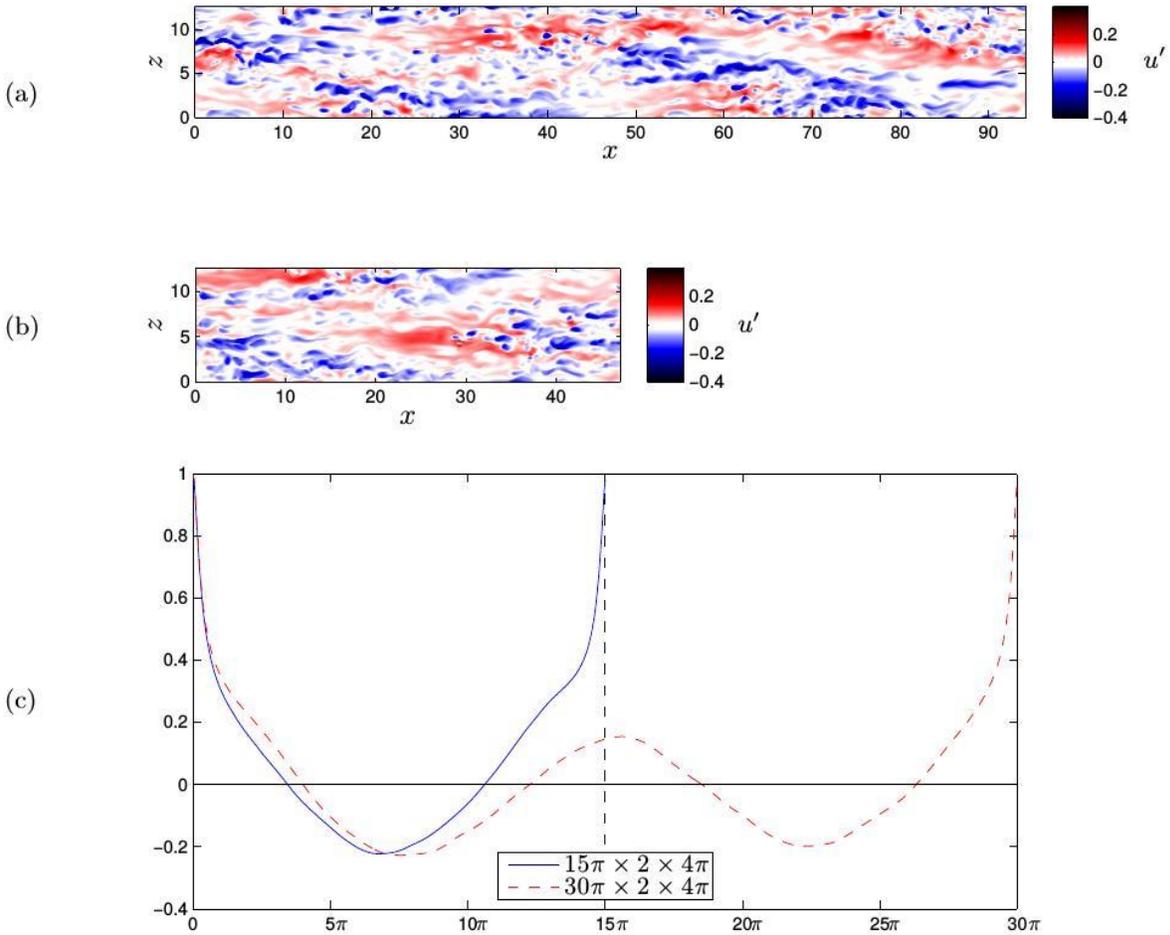

Figure 4: Adequate domain sizes; $Re_{Dh} = 4000$

(a) domain size $30\pi \times 2 \times 4\pi$

(b) domain size $15\pi \times 2 \times 4\pi$

(c) autocorrelation functions

## 2.2. Domain size variations for $Re_{Dh} = 4000$

When large scale structures occur in the infinitely extended plane channel, with only the y-direction restricted to $L_y/\delta = 2$, periodic boundary conditions obviously can be set adequately or inadequately. This category refers to the question, whether the periodic boundary conditions strongly interfere with the large scale structures or not. A first simple test with respect to this issue is to check the autocorrelation functions when the domain size is doubled in one direction. When it adequately accounts for the large structures, these are preserved after the domain size is enlarged. Otherwise the two different domain sizes both strongly interfere with the structures in their own way and no similarity of the solutions can be expected.

Figures 4 and 5 show both options. In these figures the flow field is visualized by showing the velocity fluctuations u′ in the mid-plane, i.e. for $y = 0.0$. For an adequate domain size negative values of $R_{uuX}$ occur in figure 4 which will be explained later. They are the "footprint" of quasi-stable large scale structures which obviously appear twice as often when the domain size is doubled. For an inadequate domain size in figure 5 the situation is very different, however. Neither does the autocorrelation for the small domain reach negative values nor is there a doubling of events when the domain size is increased by a factor of two. Out of all the (six) different domain sizes, which have been applied for $Re_{Dh} = 4000$, cf. figure 2, only those shown in figure 4 do not strongly interfere with the large scale structures. The friction factor values of these solutions are almost the same, whereas for all other cases they deviate by several percent, see figure 2 (b).

## 2.3. Stripe patterns

The stripe patterns which clearly appear in figure 4 can also be observed in experiments. Hashimoto et al.(2009) found such streak like agglomerations which were inclined at an angle of ≈ 25° against the streamwise direction. These structures can be tilted both to the right and to the left so that an overall situation like sketched in figure 6 exists, with two wavelength components λx , λz and α = arctan (λz /λx ). Only when the numerical solution domain fulfills the conditions

$$L_x = N\,\lambda_x \quad \text{and} \quad L_z = N\,\lambda_z \quad \text{with} \quad N = 1, 2, 3, \dots \quad (2.6)$$

it is a fully adequate solution domain reproducing the real angle $\alpha$ seen in experiments. This approximately was the case in figure 4, but not in figure 5. Since periodic boundary patterns conditions appear enforce a certain artificial streak inclination angle $\hat{\alpha} = \arctan(L_z/L_x)$ from stripe patterns appear in a numerical certain artificial solutions streak when $\hat{\alpha}$ is close to $\hat{\alpha} = \arctan(\lambda_z/\lambda_x)$ from the real physics, observed in experiments. When, however, $\hat{\alpha}$ is not close to the real $\alpha$ it may happen that no large scale structures are predicted at all, as in figure 5 (b).

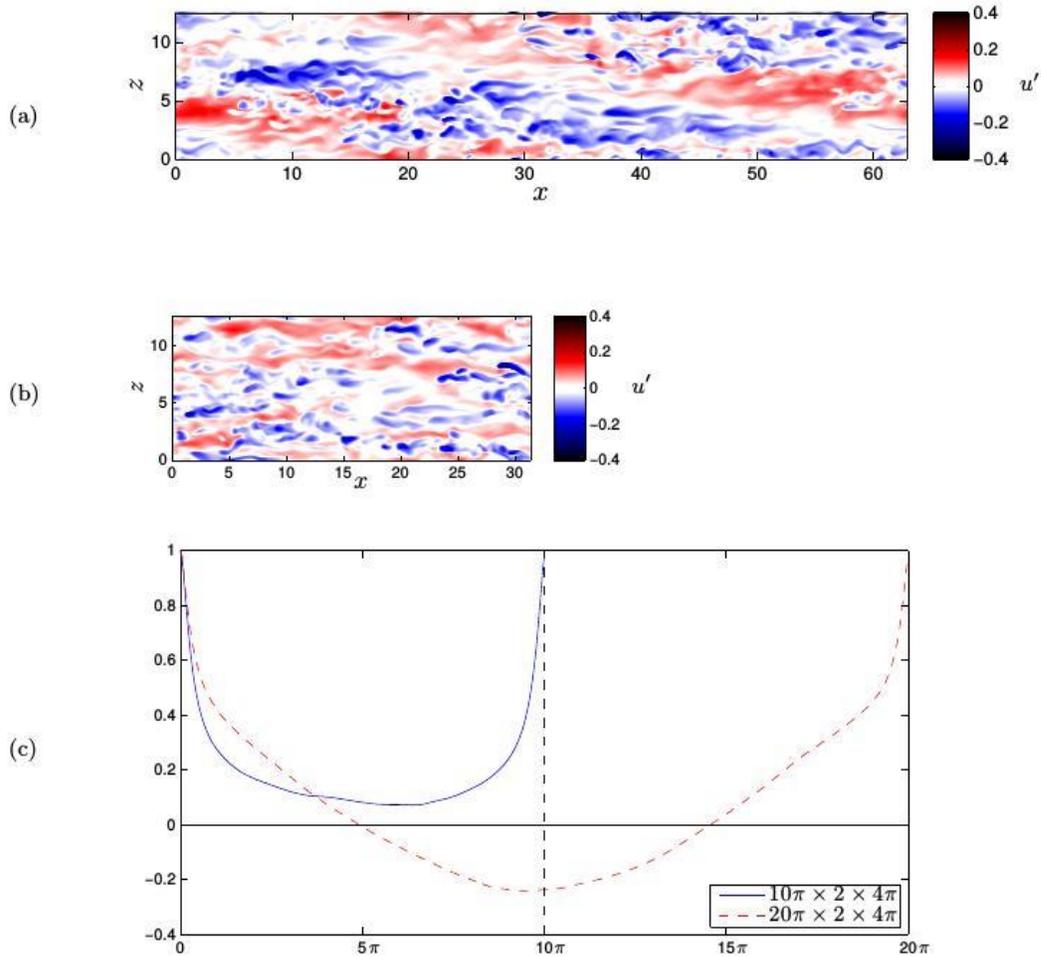

Figure 5: Inadequate domain sizes; $Re_{Dh} = 4000$

(a) domain size $20\pi \times 2 \times 4\pi$

(b) domain size $10\pi \times 2 \times 4\pi$

(c) autocorrelation functions

With these considerations in mind we select two special test cases TC-1 and TC-2 with solution domains, in which one large scale structure appears, i.e. with N = 1 in (2.6). Details are given in table 1 in terms of domain size ($L_x$, $L_y$, $L_z$) from which $\hat{\alpha}$ follows, Details number are of given grid in points table ($N_x$, $N_y$, $N_z$), and step sizes ($\Delta x^+$, $\Delta y^+|_w$, $\Delta z^+$) with $\Delta x^+ = \Delta x u_\tau/\nu$, for example. Also given are the global results $Re_\tau$, $f_w$, and the Nusselt number $Nu$.

Figure 7 shows the results for both test cases including their autocorrelation functions. They will be discussed in the following chapter.

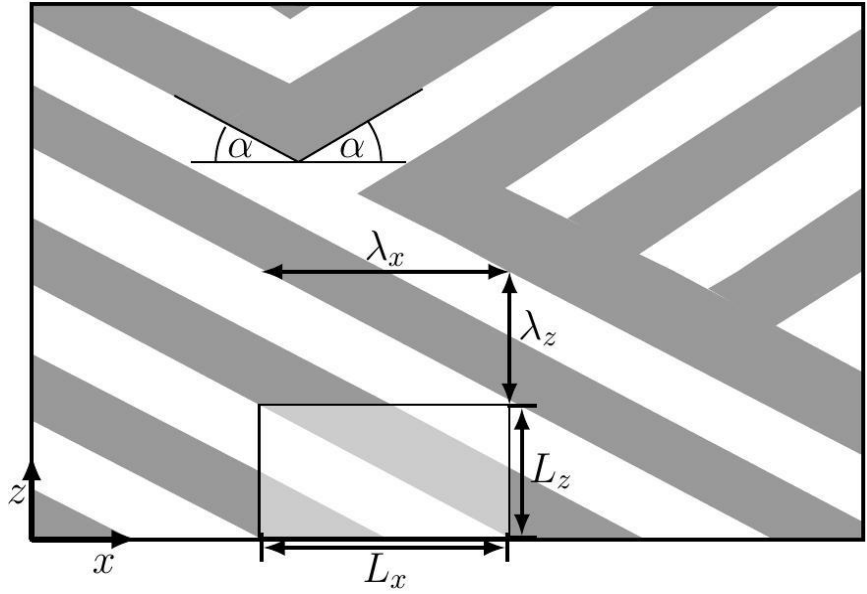

Figure 6: Streak like structures (stripe patterns) in the experiments by Hashimoto et al.(2009) and adequate minimal domain sizes $L_x$, $L_z$; $\lambda_x$, $\lambda_z$ : wave length components of streaks with laminar character, α: streak inclination angle against the streamwise direction

|  | $L_x \times L_y \times L_z$ | $\hat{\alpha}$ | $N_x \times N_y \times N_z$ | $\Delta x^+$ | $\Delta y^+|_w$ | $\Delta z^+$ | $Re_\tau$ | $f_w$ | Nu |
|---|---|---|---|---|---|---|---|---|---|
| TC-1: | $18\pi \times 2 \times 8\pi$ | 24° | $480 \times 65 \times 512$ | 8.37 | 0.086 | 3.57 | 71.1 | 0.04007 | 2.45 |
| TC-2: | $20\pi \times 2 \times 10\pi$ | 26.6° | $512 \times 65 \times 640$ | 8.67 | 0.085 | 3.47 | 70.7 | 0.03998 | 2.22 |

Table 1: Two test cases (TC) and details of the numerical grid with $Re_{Dh} = \rho u_m 4\delta/\eta = 4000$ and Pr $= c_p\eta/k = 0.71$; the bulk velocity um is enforced as a constant constraint

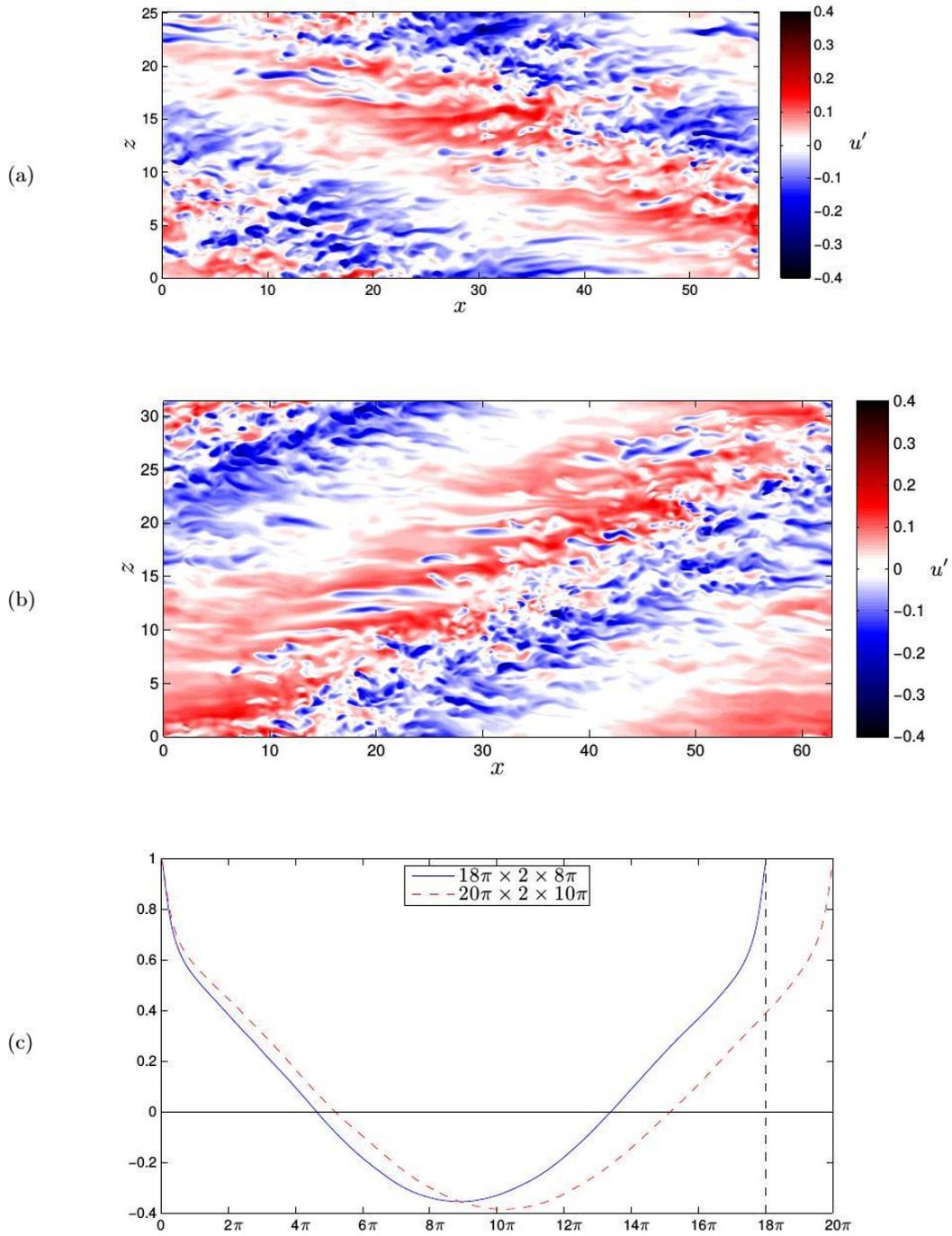

Figure 7: Test cases TC-1 and TC-2 with $\widehat{\alpha} = \arctan(L_z/L_x) \approx 25°$

(a) TC-1: $18\pi \times 2 \times 8\pi$; $\hat{\alpha} = 24°$

(b) TC-2: $20\pi \times 2 \times 10\pi$; $\hat{\alpha} = 26.6°$

(c) autocorrelation functions; $Re_{Dh} = 4000$

## 3. The physics of large scale structures in the flow field

Both test cases in figure 7 have a domain angle $\hat{\alpha} \approx 25°$ which is sufficiently close to the (initially unknown) streak inclination angle $\alpha$ so the large scale structures emerge in the solutions. Comparing the corresponding autocorrelation functions in figure 7 (c) shows, however, that a certain choice of ˆ has an impact on the solution even close to r = 0 since $\alpha \equiv \hat{\alpha}$ is enforced by this choice. This is different when no large scale structures are present since then a change in domain size does not affect the autocorrelation functions in the vicinity of their origin at $r = 0$, as can be seen in figure 3 for such a case.

Since negative values of the autocorrelation functions are the "footprint" of large scale structures one might argue that cases with lower (negative) values of $R_{uuX}$, like TC-2 compared to TC-1 in figure 7 (c), are those with the more adequate solution domain. This, however, assumes that large scale structures cannot be unphysically reinforced by certain (inadequate) domain sizes. A reliable answer to this open question might be found from solution domain sizes with $N \gg 1$ in (2.6), so that the characteristic length of the large structures would be small compared to that of the solution domain. Tsukahara & Kawamura (2007) performed DNS calculations with a domain size they called "huge box" ($328 \times 2 \times 128$). They found, based on the energy spectrum for $u'$, that the most energetic wave lengths occur with $\alpha \approx 21.8°$, which is very close to their $\hat{\alpha} \equiv \arctan(L_z/L_x) \approx 21.3°$, so that even for the "huge boxes" there may be a triggering effect on the large scale structures.

### 3.1. *"Footprints" of large scale structures*

With a domain size like indicated in figure 6, i.e. with a length $L_x$ that covers one element of the stripe pattern, an autocorrelation $L_x/2$ appart can be expected to be non-zero when stripes occur. Then the two correlated locations lie predominantly in different parts of the stripe patterns (grey and white in figure 6).

For such a case in figure 8 velocities at two locations in the midplane y = 0 of the channel are shown for a time period in which more than one single stripe passes through the solution domain (here $\Delta t = 100$ with $\Delta t \approx 60$ for a single stripe to pass by). These velocities fluctuate around the time averaged value u in a specific manner and not in a completely random way. There are periods (marked grey in figure 8; width $\approx \Delta t = 30$, i.e. one half of a stripe pass time) in which one velocity is well above $\bar{u}$ and rather smooth with respect to its time dependence while the other velocity is well below u and strongly fluctuating. This happens alternatingly and can be attributed to the stripe patterns moving with the flow and thus through the solution domain. Since these deviations from u appear regularly and with a different sign their contribution to $R_{uu,X}$ lets the autocorrelation function become negative around $r = L_x/2$, what was named the "footprint" of large scale structures.

It should be noted that velocities become smoother with respect to their time dependence for $u > \bar{u}$ and increase their fluctuations for $u < \bar{u}$. Due to the nondimensionaliza- tion of the velocity with $u_c = 1.5$um (centerline velocity), a completely laminar flow would have $u = 1$ in the midplane while $u = 2/3$ would be the value for a fully turbulent flow at Re = ∞. Thus, velocities above $\bar{u}$ tend to u = 1 and those below $\bar{u}$ to u = 2/3 which corresponds to a predominantly laminar or turbulent character, respectively, visible in the plots for u′ for example, see figure 7 (a) and (b).

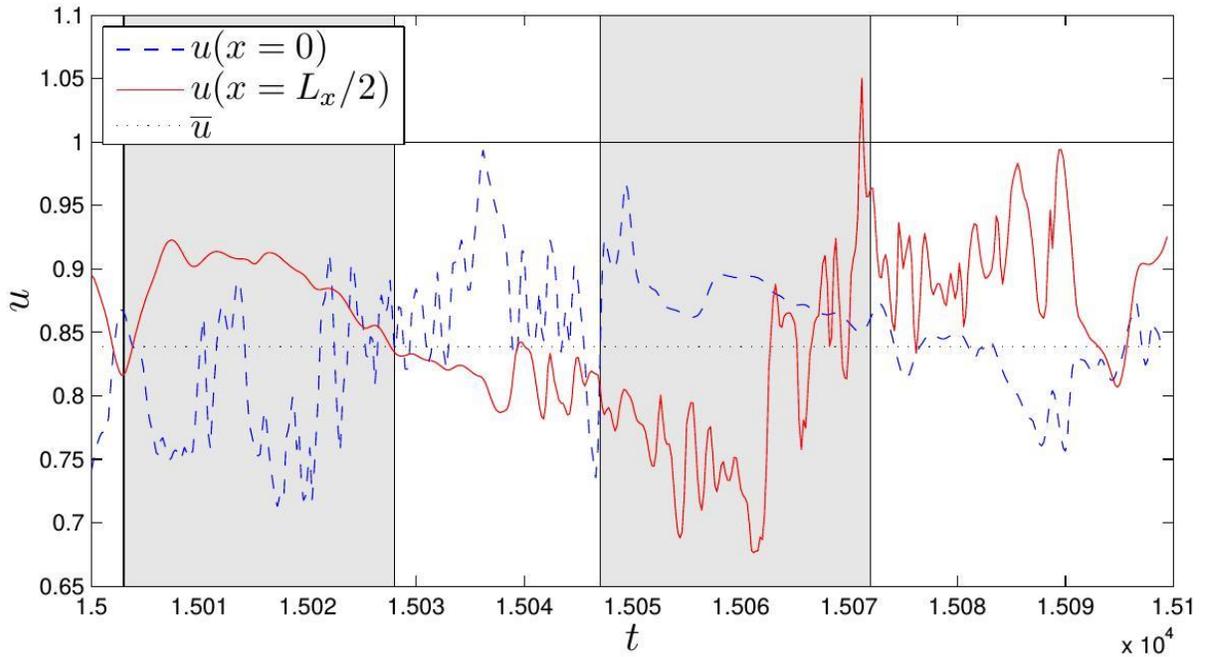

Figure 8: Time dependent velocities at two locations in the midplane $L_x/2$ apart; test case TC-2

### 3.2. *Large scale structures and their effect on the mean flow*

Turbulent large scale structures appear at Reynolds numbers well below the critical Reynolds number $Re_{Dh} = 3848$ according to the linear stability theory, see Schäfer & Herwig(1993). While this finite number exists for plane channel flows the laminar base flow is linearly stable for all Reynolds numbers for circular duct and Couette flows. Since, however, the large scale structures for Couette flows, for example, are similar to those of the plane channel flow their origin may be due to local instabilities but not specifically due to the physics of linear instability growth in a certain geometry. There are various explanations for the origin of the stripe patterns found in a certain small band of Reynolds numbers, reaching from a scenario "noise → streaks → spot nucleation → stripe" (Duguet et al.(2009)) to "growth by destabilization" (Dauchot & Daviaud(1995)).

Whatever the exact mechanism is, it leads to distinct and stable structures that can be observed in DNS solutions as well as in experiments. These stable structures and their time averaged statistics (and not their formation) are in the scope of our study, based on DNS results in a long time period. The length of this time period is determined by our criterion that friction factors $f_p$, $f_w$ and $f_\phi$ according to (2.1) and (2.2) may not differ by more than 0.2% (c.f. appendix). In order to reach time averaged values which then no longer depend on the time of averaging, more than 120 dimensionless time units are required here, i.e. a passive scalar must travel more than 120 times through the solution domain in streamwise direction at a constant speed $u_m$.

For an inspection of the instantaneous flow field the so–called $\lambda_2$–method is applied for visualizing details of the vortical structures at a certain instant of time. Within this method pressure minima are identified while taking into account of unsteady strain and viscous effects. Hence, according to Jeong & Hussain (1995), a vortex corresponds to a region where two eigenvalues of the symmetric tensor $\mathbf{S}^2 + \mathbf{\Omega}^2$ are negative; here $\mathbf{S}$ and $\mathbf{\Omega}$ are the symmetric and antisymmetric parts of the velocity gradient tensor $\nabla \mathbf{u}$.

Figure 9 shows iso-surfaces with $\lambda_2 = -0.01$ which reveal the vortex substructure of the stripe patterns. Most of the elongated substructures are close to the walls and inclined away from the wall by an angle $< 30°$. However, also single structures with inclination angles $\approx 45°$ appear, as indicated in figure 9 b). For wall bounded flows they are known as hair pin vortices and play an important role in turbulence production, see Moin & Mahesh (1998).

As far as the stripes as a whole are concerned there are two important aspects with respect to their motion in the flow field, which are their speeds in *x*- and *z*-direction, respectively. Often the speed in x-direction is not considered further stating that "the pattern moves at a constant velocity, which is almost (the; added here) same with the bulk mean velocity" (Tsukahara & Kawamura(2007)). This is true, however, only as a first approximation. A closer look at the moving stripe pattern reveals that neither its horizontal streamwise velocity is that of the bulk mean velocity nor that its lateral velocity is zero.

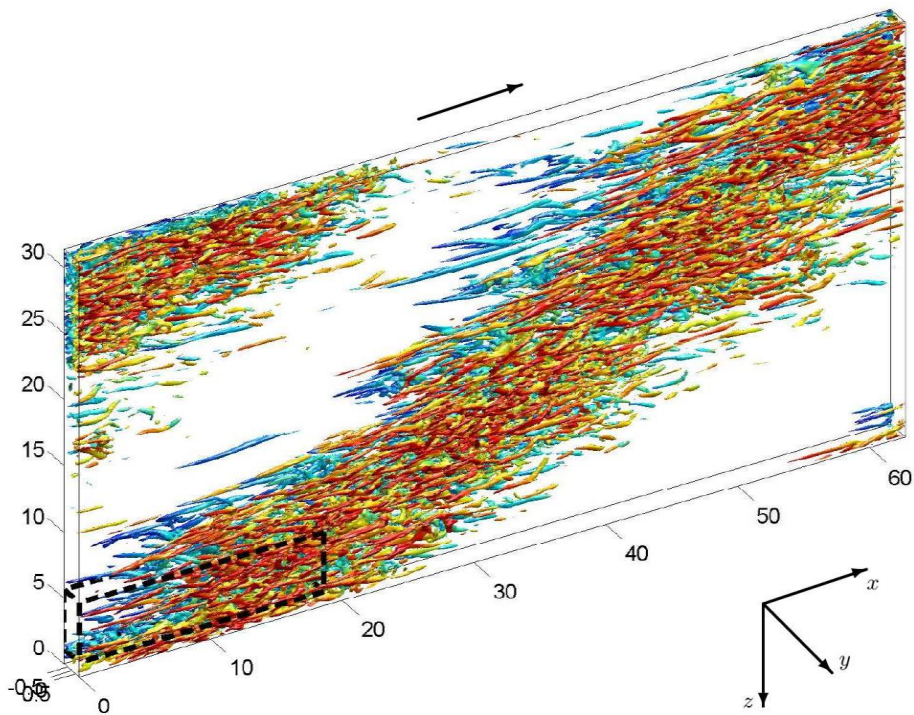

(a) Flow details in the whole solution domain

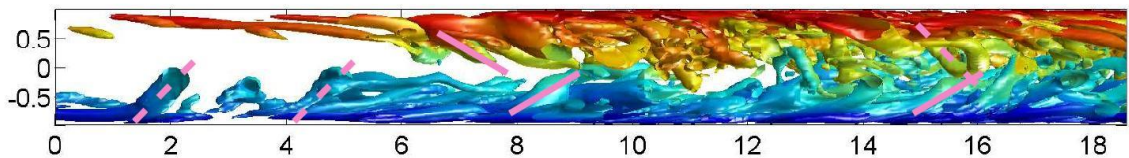

(b) top view on the box marked in the lower left corner in (a), inclinations: $-30°$; $--45°$

Figure 9: Inner structure of a turbulent stripe of test case TC-2 visualized by the $\lambda_2$-method

For a reasonable analysis of the stripe pattern motion a characteristic velocity of the pattern as a whole should be defined. Since the stripe consists of changing agglomerations of certain parts of the moving fluid one can only find a characteristic velocity which is physically reasonable but still to a certain extend arbitrary. With this in mind we define a stripe pattern velocity us as following, demonstrated for the TC-2 case.

First the velocity component $u(x)$ at a time $t$ and at a certain position $(y = 0, z_i)$ is approximated by the Fourier series expansion.

$$u(x) = \frac{a_0(z_i)}{2} + A_1(z_i)\cos\left(\frac{2\pi}{L_x}x - \varphi_1(z_i)\right) + \sum_{n=2}^{N_x/2} A_n(z_i)\cos\left(\frac{2\pi n}{L_x}x - \varphi_n(z_i)\right) \quad (3.1)$$

If just one stripe occurs within the solution domain as in TC-2, then the $n = 1$ term will account for the general structure of stripe patterns as alternating regions of higher and lower streamwise velocities u. In equation (3.1) $\varphi_1(z_i)$ is the phase shift of the $n = 1$ term which will be different for different z-positions since the stripe pattern is tilted by a certain angle. In order to determine these phase shift values for all discrete zi-positions on our grid simultaneously we assume a fixed angle of arctan ($L_z/L_x = 0.5$) for the tilt of the stripe pattern. Due to the periodic boundary conditions this value is exact in the time averaged sense. Hence, the combined phase shift (averaged over all zi-positions) assuming a fixed tilt angle of the stripe pattern with $z_i = L_z n_z/N_z$ is

$$\widetilde{\varphi_1} = \sum_{n_z=0}^{N_z-1}\left[2\pi\frac{n_z}{N_z} + \varphi_1\left(L_z\frac{n_z}{N_z}\right)\right] \quad (3.2)$$

For one wash out cycle the combined phase shift ~ will vary strictly monotonically within [0; 2π] as the stripe pattern moves through the solution domain. Therefore we introduce a characteristic velocity us in terms of the combined phase shift velocity.

$$u_s(t) = \frac{L_x}{2\pi}\left.\frac{\partial\widetilde{\varphi_1}}{\partial t}\right|_t \approx 10\frac{\Delta\widetilde{\varphi_1}(t)}{\Delta t} \quad (3.3)$$

The result of 31 wash out cycles of the stripe pattern for TC-2 is shown in figure 10. It turns out that the stripe pattern is slightly faster than the bulk mean velocity ($u_s/u_m$ = 1.028). Conversely that indicates that the stripe pattern is much slower than the mean centerline velocity uc (mean value of $u_s/u_c$ = 0.817); i.e. $u_m < u_s \ll u_c$. However, the velocity of the stripe pattern us is not constant but fluctuates rapidly. This volatility is the result of the stripe forming mechanism itself: production of turbulent kinetic energy prevails in front of the predominantly turbulent region whereas it is dissipation on its trailing edge. Hence, volatility of the production and dissipation rates results in a thinner, wider, slower or faster stripe pattern.

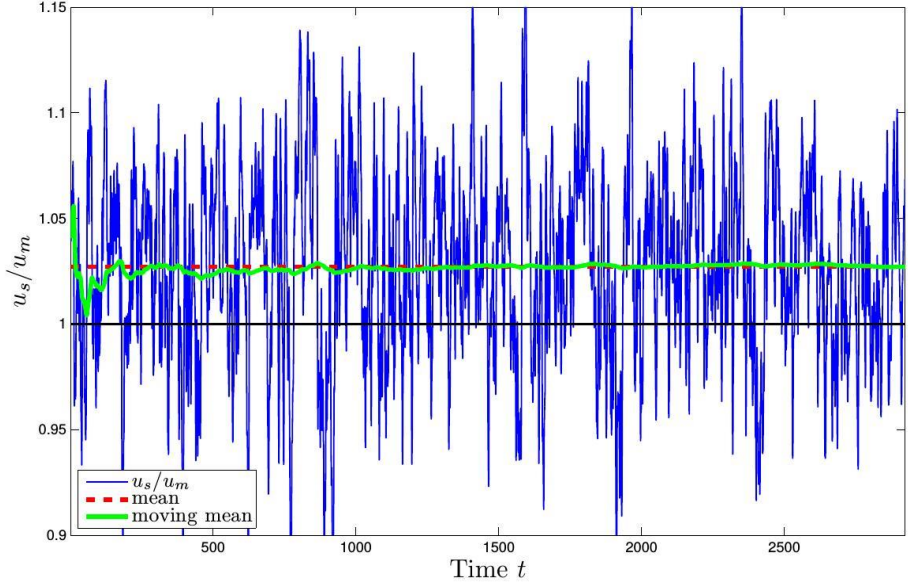

Figure 10: Development of the stripe pattern velocity us together with its moving mean and mean after 31 wash out cycles

The stripe pattern is not only characterized by its streamwise velocity us but also by a mean lateral velocity as already stated above. A careful inspection of the DNS results as a function of time reveals, that between the predominantly laminar parts of the overall structure and those parts with strong turbulence there is a slowly gliding motion relative to each other as indicated in figure 11. As a consequence there is a certain (locally averaged) velocity component in $z$–direction, which in the more laminar part of the stripe pattern will be positive and which will be negative in the highly turbulent part. Then there will be wall shear stress components $\tau_{yz,w}$ in $z$–direction with $\tau_{yz,w} > 0$ in the laminar and $\tau_{yz,w} < 0$ in the turbulent part of the stripe patterns. Since, however, there is no pressure gradient in z–direction, the overall, time averaged wall shear stress in $z$–direction must be zero. From these considerations we conclude that in the near wall region a larger fraction of the flow will be in the laminar part since corresponding laminar shear stresses are smaller than those resulting from turbulent flows.

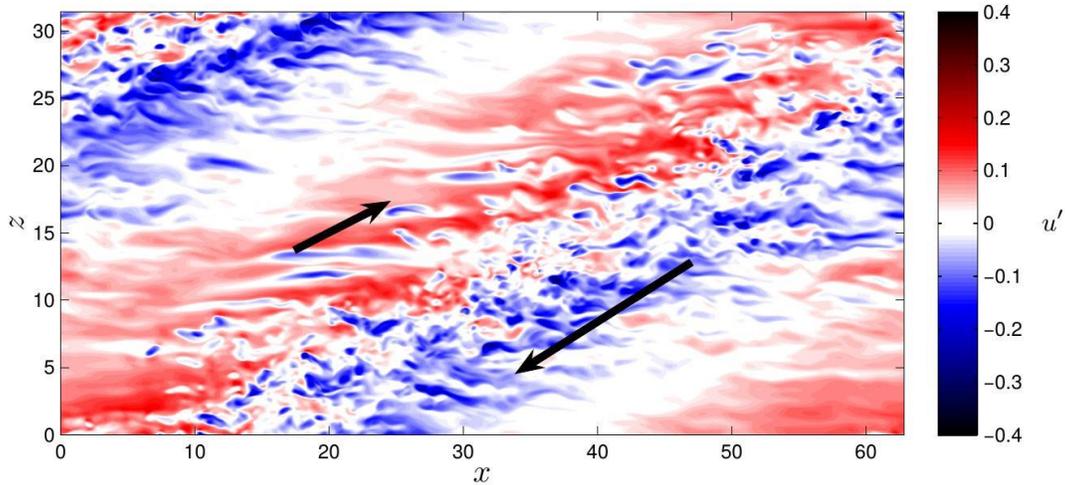

(a)

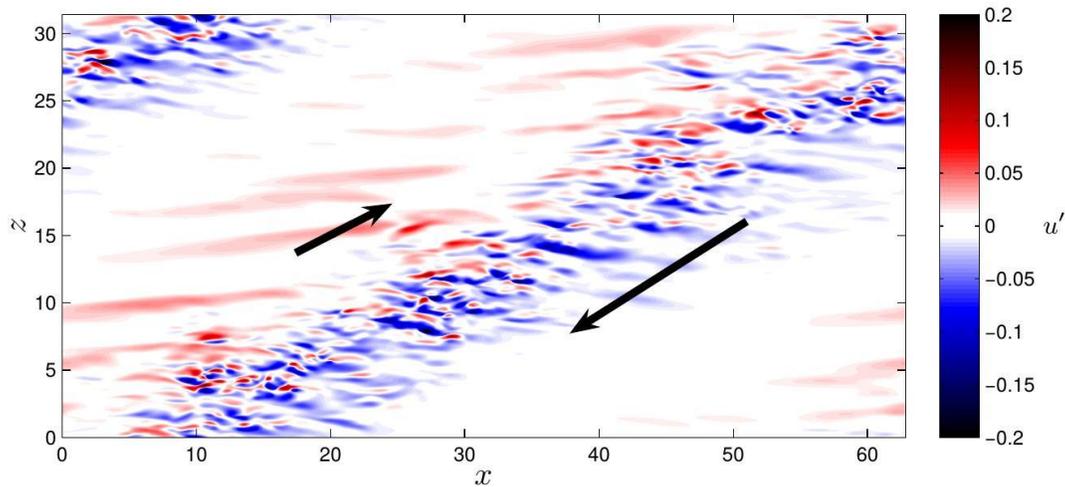

(b)

Figure 11: Sketch of the gliding motion within the stripe pattern of test case TC-2

(a) at the mid-plane of the channel, color coding represents u-velocity fluctuations with respect to the mean flow in streamwise direction

(b) in the vicinity of the wall ($y = 0.043$), color coding represents w-velocity (spanwise direction)

This is confirmed by figure 11(b) since the increase of the velocity in z-direction w, corresponds directly to the local wall shear stress, $\tau_{yz,w} \approx \Delta w/\Delta y$. Compared to figure 11(a) the extent of the turbulent region is reduced due to damping of the wall although the local magnitude of $\tau_{yz,w}$ is still higher in the turbulent part. Therefore, the net mass flux in $z$-direction and thus the pressure gradient is zero close to the wall, since both distribution and magnitude of $\tau_{yz,w}$ compensate each other ($\int \tau_{yz,w} dA = 0$). In the center region, however, the laminar and turbulent regions are equally distributed and since the negative w component in the turbulent region dominates (as indicated by the two arrows) the overall mass flux in the center of the channel is negative. Such a velocity component indeed occurs as shown in figure 12. In addition to the u–component in x–direction there is a *w*–component, which has zero wall gradient ($\to \tau_{yz,w} = 0$), a maximum value |w| ≈ 2.9% of u and a mean value $w_m$ ≈ 1.8% of $u_m$.

This $\bar{w}$–component is a consequence of the gliding motion between the "laminar" and the "turbulent" parts of the stripe patterns. So far it remains an open question whether a finite lateral extension of the channel in which there can be no such overall lateral motion would lead to a suppression of the structures as a whole or would just result in a modification of it.

## 4. Entropy generation in the flow and temperature field

From a thermodynamics point of view entropy generation in a flow field indicates losses of exergy (also called available work) and thus is immediately related to the dissipation $\varphi$ and, as a whole to the friction factor $f_\varphi$ according to (2.2). In the temperature field, entropy generation (due to heat conduction in a non-isothermal temperature field $\theta$; c.f. table 1 for Nusselt numbers) indicates an irreversible heat transfer with loss in quality of the transferred energy, see Herwig (2011) for more details with respect to the role of entropy in heat transfer processes.

Both kinds of entropy generation can be identified in the channel flow under consideration here and helps to understand the physics of momentum and heat transfer processes. As an example we assumed a non-dimensional temperature difference $\varepsilon = \Delta T/T_m$ of 0.01. The choice of $\varepsilon$ is quite arbitrary as it basically is a thermal boundary condition. Hence, $\varepsilon$ is a tuning parameter for our post processing, nonetheless it is crucial for the strength of entropy generation (normalized by $k/\delta^2$) which will be shown in the following.

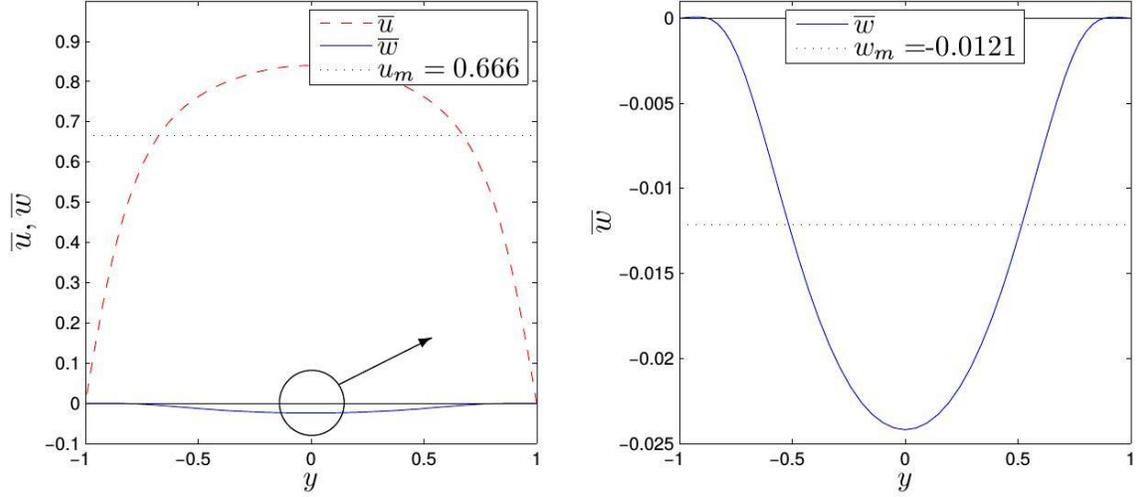

Figure 12: Time averaged velocity components u(y) and w(y) of test case TC-2

### 4.1. *Entropy generation in the flow field*

According to the second law of thermodynamics entropy generation occurs in a flow field when there are local and instantaneous velocity gradients. The instantaneous and local non-dimensional entropy generation rate $\dot{S}_D'''$ is, cf. Herwig & Wenterodt(2011)

$$\dot{S}_D''' = \text{Ec}\,\text{Pr}\,\frac{\varepsilon}{\varepsilon\theta + 1}\,\frac{1}{2}\left(\frac{\partial u_i}{\partial x_j} + \frac{\partial u_j}{\partial x_i}\right)^2 \qquad (4.1)$$

For the sake of simplicity we assumed Ec Pr = 1, with Ec = $u_{ref}^2/c_p\Delta T$. While we had to choose the Prandtl number beforehand (Pr=0.71) as it appears in the thermal energy equation, the Eckart number is just like $\varepsilon$ a tuning parameter for our post processing. Note that for $\Delta T \to 0$ and therefore $\varepsilon \to 0$ the entropy generation rate remains finite, since Ec $\varepsilon = u_{ref}^2/c_p T_m$ is unchanged. Figure 13 shows the instantaneous distribution of the entropy generation due to dissipation given by (4.1) for three different but constant values of $\dot{S}_D'''$. For a better resolution only the lower left quarter of the flow field of figure 9 is shown here. Since these values correspond to the strength of dissipation, high values will be found predominantly close to the wall in regions of laminar character while they will spread over the whole channel width when there is a strong turbulent motion. This can be seen in figure 13 (a)-(c) and more pronounced in figure 14 which shows details in the upper left corner of the domain of figure 13 (a). This corner is in the middle of the turbulent stripe and reveals the finer structure of the flow field.

Altogether this structure of the entropy generation field is in complete accordance with the interpretation of the stripe structure given so far.

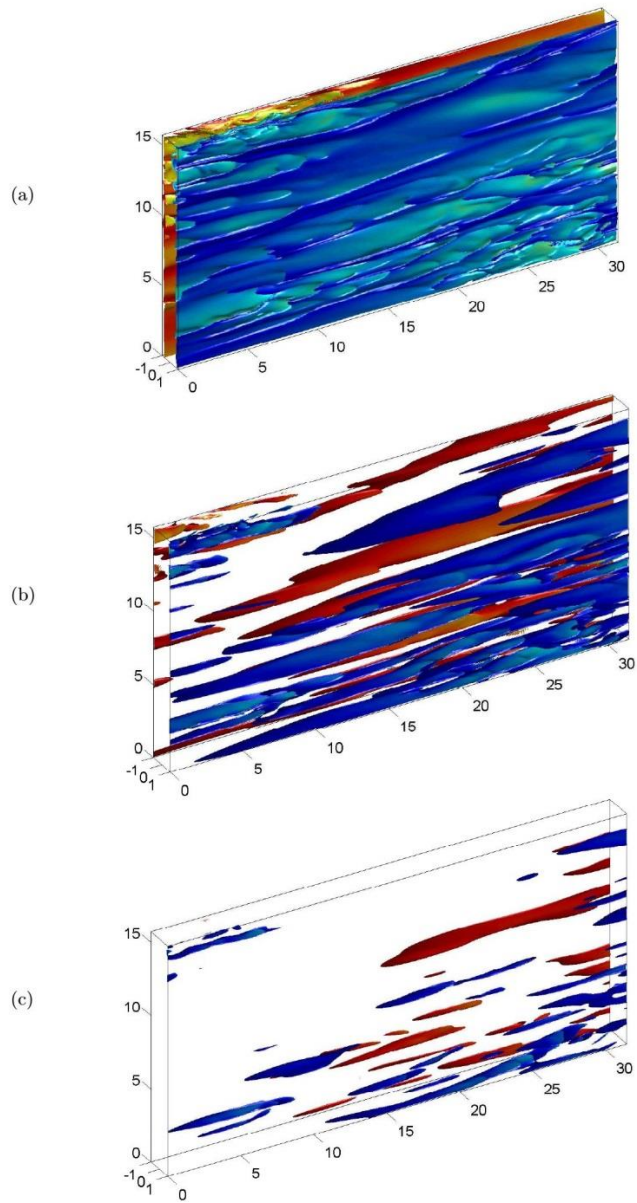

Figure 13: Entropy generation due to dissipation. The color coding is according to the temperature, blue: cold wall in front, red: hot back wall.

(a) $\dot{S}_D''''=2$, (b) $\dot{S}_D''''=10$, (c) $\dot{S}_D''''=22$

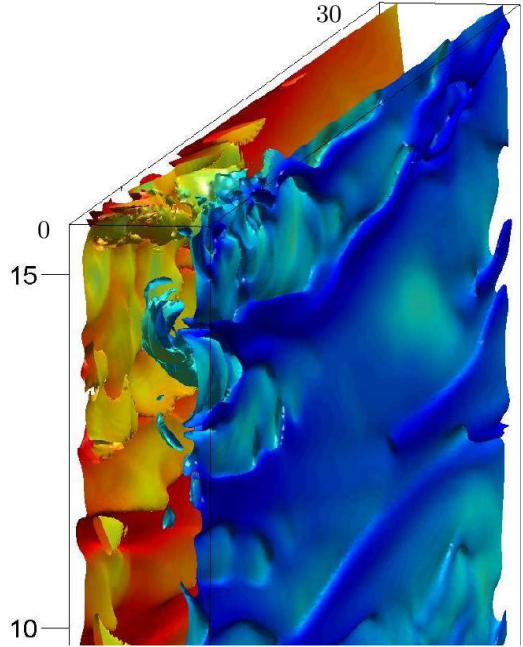

Figure 14: Details of the entropy generation field shown in figure 13. Here: $\dot{S}_D''' = 2$

4.2. *Entropy generation in the temperature field*

According to the second law of thermodynamics entropy generation occurs in a temper- ature field when there are local and instantaneous temperature gradients. The instantaneous and local non-dimensional entropy generation rate $\dot{S}'''$ is, cf. Herwig & Wenterodt (2011)

$$\dot{S}_C''' = \left(\frac{\varepsilon}{\varepsilon\theta+1}\right)^2 \left(\frac{\partial \theta}{\partial x_i}\right)^2 \qquad (4.2)$$

Here, this entropy generation in the temperature field is due to a heat transfer caused by a temperature difference $\Delta T = T_h - T_c$ between the hot and the cold wall. In figure 15 again three constant values, now of $\dot{S}_D'''$, are shown in the same part of the flow field as in figure 13. From this one can see basically the same structures as in the flow field, indicating that the temperature acts as a passive scalar in the dominating flow field. In terms of turbulence modelling this justifies the often used close link of an eddy thermal conductivity to the eddy diffusivity via a (constant) turbulent Prandtl number. As a detail in the turbulence dominated region figure 16 again shows the upper left corner in the flow field shown in figure 15.

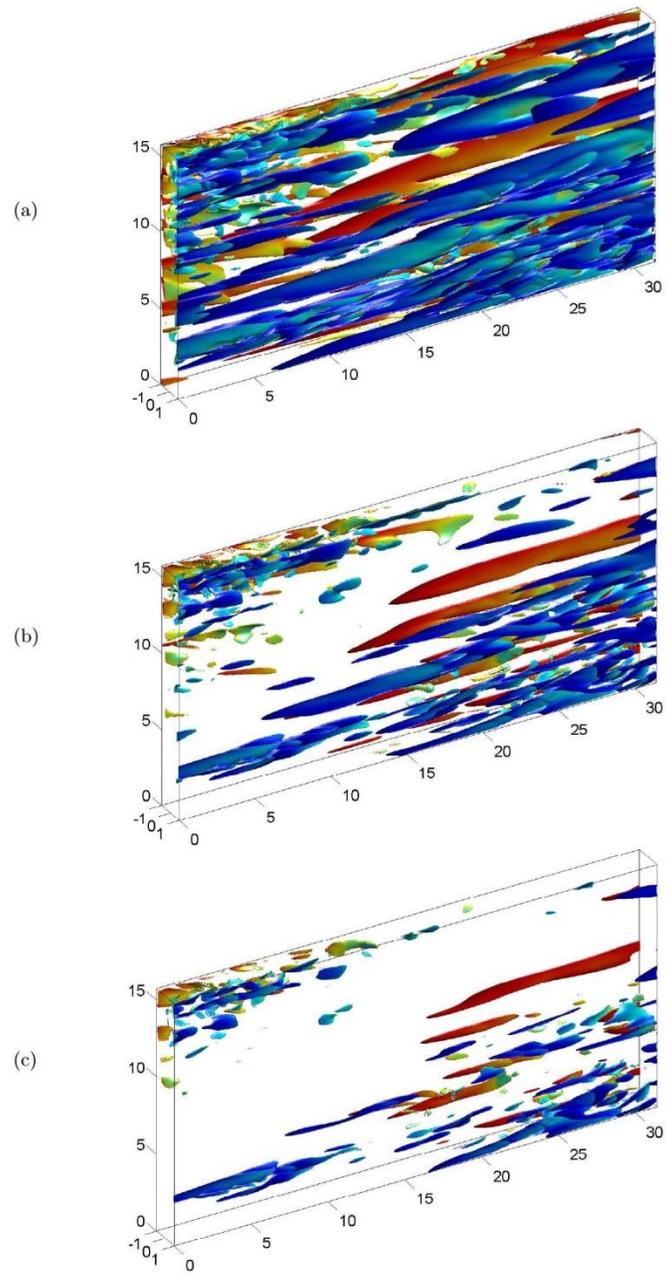

Figure 15: Entropy generation due to heat conduction. Color coding like in figure 13.
(a) $\dot{S}'''_C=1$, (b) $\dot{S}'''_C=2$, (c) $\dot{S}'''_C=3$

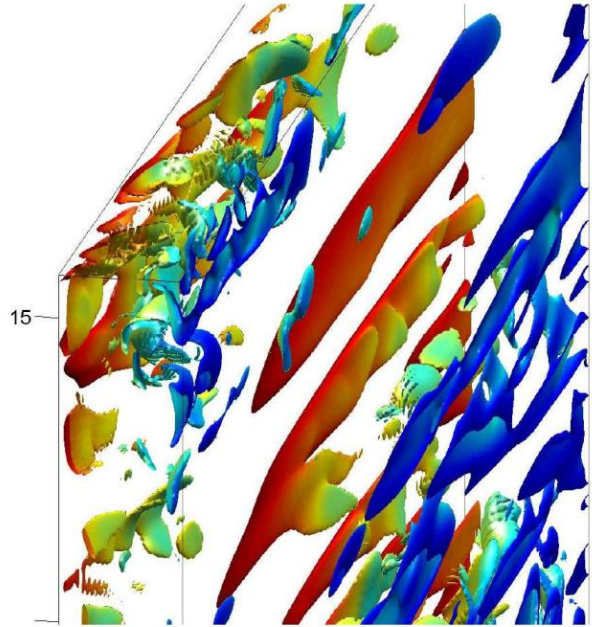

Figure 16: Details of the entropy generation field shown in figure 15. Here: $\dot{S}_C''' = 2$

## 5. Conclusion

DNS results in finite parts of an infinitely extended plane channel have been determined for Reynolds numbers well above the critical Reynolds number but low enough for laminar regions to exist in the flow field. Laminar/turbulent stripe patterns appear when the aspect ratio of the finite solution domain is carefully adopted to the physics prevailing in an infinitely extended flow domain. Time mean values can be determined by turbulent statistics in sufficiently long time periods of the DNS solutions. Special aspects of our study to reveal the physics of these flows are:

- the careful choice of the aspect ratio of the finite solution domain, c.f. figure 4 and 5
- the impact of periodic boundary conditions on the autocorrelation function, c.f. figure 3
- the "footprint" of large scale structures as negative values of the autocorrelation function, c.f. figure 7
- the occurrence of a mean axial velocity of the stripe pattern us with $u_m < u_s \ll u_c$, c.f. figure 10

- the occurrence of a lateral mean flow velocity without lateral pressure gradient, c.f. figure 12
- Friction factors and Nusselt numbers determined from the corresponding entropy generation rates, c.f. table 1

Altogether from the DNS results and their careful interpretation one obtains a precise and clear picture of the complex flow and heat transfer physics in this absolutely simple geometry.

## 6. Appendix

For our computations, the nondimensionalised incompressible Navier-Stokes equations together with the thermal energy balance are integrated in a velocity-pressure formula- tion using a third-order semi-implicit Adams-Bashforth/Backward Differentiation scheme (ABBDI3). The fluid is assumed to be Newtonian and the temperature acts as a passive scalar since the momentum equation is decoupled from the thermal energy balance. The equations in Einstein notation are

$$\frac{\partial u_i}{\partial x_i} = 0 \tag{6.1}$$

$$\frac{\partial u_i}{\partial t} + u_i \frac{\partial u_j}{\partial x_i} = -\frac{\partial p}{\partial x_i} + K_u \frac{\partial^2 u_i}{\partial x_j^2} \tag{6.2}$$

$$\frac{\partial \theta}{\partial t} + u_i \frac{\partial \theta}{\partial x_i} = K_\theta \frac{\partial^2 \theta}{\partial x_j^2} \tag{6.3}$$

With the physical properties at the mean temperature between the hot and the cold walls, $T_m^* = (T_h^* + T_c^*)/2$, the temperature difference $\Delta T^* = T_H^* - T_c^*$ and the channel half width $\delta^*$ as reference quantities, the nondimensional parameters $K_i$ only depend on the choice of the reference velocity $u_{ref}^*$. They are

$$K_u = \frac{\nu^*}{u_{ref}^* \delta^*} \qquad K_\theta = K_u/\text{Pr} \tag{6.4}$$

The reference velocity $u_{ref}^*$ was chosen as the centerline velocity of a laminar flow $u_{c,lam}^* = 3u_m/2$ due to the special nature of the flow under consideration. Hence, if the flow becomes laminar the nondimensional centerline velocity is $u_c = u(y = 0) = 1$. If the flow becomes fully turbulent (Re $\to \infty$) the centerline velocity becomes uc $= u_c = u_m^* / u_{c,lam}^* = 2/3$.

The spatial discretization is based on a pseudospectral Chebychev-tau method (for details see Peyret(2002)) with periodic boundary conditions using Fourier series in both streamwise (*x*) and spanwise (*z*) directions. The nonlinear terms of the momentum equation are treated explicitly in a vorticity-velocity representation, while aliasing is prevented by applying the 3/2 rule in every direction. The pressure field is solved directly utilizing the influence-matrix following Kleiser & Schumann(1980) and the tau error arising from the no-slip condition at the walls is corrected. The mean pressure is adjusted dynamically to maintain a constant mass flux ($u_m = u_m^* / u_{ref}^* = 2/3$). The mechanical energy equation is derived by multiplying the momentum equation (6.2) with the velocity vector ui. This equation can be time averaged and integrated over the entire domain using the numerical periodicity in the streamwise and spanwise directions, resulting in

$$\underbrace{-u_m \frac{\partial \bar{p}}{\partial x}}_{f_p u_m^3/8} = \underbrace{K_u \int_{-1}^{1} \overline{\frac{1}{2}\left(\frac{\partial u_i}{\partial x_j} + \frac{\partial u_j}{\partial x_i}\right)^2} dy}_{f_\varphi u_m^3/8} \qquad (6.5)$$

Thus, the difference between both formulations of a friction factor fp and fϕ is nothing but the residual of the time averaged momentum equation. Our computations presented in this paper obey residuals of the momentum equation of less than $2 \times 10^{-4}$, i.e. $f_p$ and $f_\varphi$ deviate by less than 0.2%.